# Contested Logistics: Resilience of Strategic Highways and Railways

**Authors**:

Sukhwan Chung

Credere Associates, LLC
776 Main St, Westbrook, ME 04092, USA

Email: sukhwan.chung@usace.army.mil

ORCiD: 0009-0004-3873-8948

Daniel Sardak

Department of Computer Science, Worcester Polytechnic Institute
100 Institute Rd, Worcester, MA, 01609-2280, USA

Email: dansardak@gmail.com

ORCiD: 0009-0006-3110-7400

Maksim Kitsak

Delft University of Technology, Delft, 2628CD, South Holland, the Netherlands

Email: m.a.kitsak@tudelft.nl

Andrew Jin

Risk and Decision Science Team, US Army Engineer Research and Development Center – Environmental Laboratory, US Army Corps of Engineers
696 Virginia Rd, Concord, MA, 01742-2718, USA

Email: Andrew.S.Jin@usace.army.mil

ORCiD: 0000-0003-1026-9217

Igor Linkov

Risk and Decision Science Team, US Army Engineer Research and Development Center – Environmental Laboratory, US Army Corps of Engineers
696 Virginia Rd, Concord, MA, 01742-2718, USA

Email: igor.linkov@usace.army.mil

ORCiD: 0000-0002-0823-8107

## Abstract

Military logistics rely heavily on public infrastructure, such as highways and railways, to transport troops, equipment, and supplies, linking critical installations through the Department of Defense's Strategic Highway Network and Strategic Rail Corridor Network. However, these networks are vulnerable to disruptions that can jeopardize operational readiness, particularly in contested environments where adversaries employ non-traditional threats to disrupt logistics, even within the homeland. This paper presents a contested logistics model that utilizes network science and Geographic Information System (GIS) to evaluate the robustness and resilience of strategic transportation networks under various disruption scenarios. By integrating GIS data to model logistics networks, simulating disruptions, and quantifying their impacts, we identified vulnerabilities in US power projection routes and assessed the resilience and robustness of highways and railways. Our findings reveal that highways are more resilient than railways, with greater capacity to absorb targeted disruptions. These findings underscore the importance of prioritizing investments in highway infrastructure and reinforcing vulnerable road and rail segments, particularly in high-risk regions, to enhance the resilience of military logistics and maintain operational effectiveness in contested conditions.

# 1. Introduction

The flow of goods in the global economy is critical to the effective deployment of humanitarian aid and military resources. The global supply chain, while increasingly intricate and interconnected, is simultaneously more vulnerable to system shocks. Recent events have starkly demonstrated how even seemingly minor delays and disruptions can trigger cascading failures, resulting in substantial economic losses, supply shortages, and, in some cases, tragic consequences involving the loss of life (Aratani, 2024; Baldwin et al., 2023; Li et al., 2024). The COVID-19 pandemic (Chowdhury et al., 2021; Dirzka and Acciaro, 2022; Li et al., 2024) and the blockage of the Suez Canal (Qu et al., 2024; Tran et al., 2025) serve as prime examples of how unforeseen events can cripple critical supply lines. These largely unintentional events had large cascading impacts that could have similarly, if not more disruptive, by intentional sabotage by adversarial actors.

While competition is inherent, and often encouraged, within commercial logistics sectors seeking market advantage through efficiency gains in cost, speed, or quality, contested logistics environments introduce a fundamentally different set of challenges. In these environments, adversaries deliberately engage in denial operations specifically designed to degrade or disable logistics capabilities. Such actions may include physical attacks, cyber warfare, and information operations targeting key logistical infrastructure. Furthermore, adversaries can exploit terrain, employ indirect fires, or deploy obstacles to upset, disrupt, and destroy supply lines. This introduces a unique set of challenges, compounded by the limited real-world examples of peer or near-peer contested logistics scenarios upon which to draw.

Recent conflicts have highlighted the difficulties of managing logistics under contested conditions. The Russo-Ukrainian War of 2022, for instance, has underscored the importance of understanding how logistics systems respond to sustained conflict and persistent disruption. Long-term strategic bombing campaigns have devastated Ukraine's energy infrastructure, particularly its power transmission network. Key supply chain issues and sustained attacks on logistical hubs have hampered the movement of goods, demonstrating how chronic degradation, rather than singular catastrophic events, can undermine logistical resilience over time (Elliott, 2024; Loyd, 2024).

The definition of "contested logistics" itself is contested. What constitutes the transition from conventional competition to a fully contested environment is not always clear. This ambiguity underscores the need for a more nuanced and comprehensive understanding of the challenges involved. As defined in 10 U.S. Code § 2926 (h), "contested logistics environment" is an environment in which the armed forces engage in conflict with an adversary that presents challenges in all domains and directly targets logistics operation. Others offer a broader interpretation: "logistics that occur under conditions wherein an adversary or competitor deliberately seeks or has sought to deny, disrupt, destroy, or defeat friendly force logistics operations, facilities, and activities across any of the multiple domains." (King, 2024). This expanded view recognizes that contested logistics can encompass a wide range of scenarios, from conventional warfare to asymmetric conflicts and even cyberattacks on critical infrastructure.

Academic literature to characterize the contested logistics environment have largely focused on strategic studies through historical contests. Quantitative modeling of contested logistics, however, remains limited. For instance, (Černý et al., 2024) developed a game-theoretic methodology to analyze contested logistics with multiple modes of transport and goods, storage nodes, and different demand profiles that could be solved with best-response mixed-integer linear programming optimizations. (Dougherty et al., 2020) performed an agent-based Monte Carlo simulation to understand optimal strategy in sea-based logistics missions where potential adversaries could detect and eliminate vessels as they traversed throughout the network. (Sorenson, 2020) used network models to understand the resilience of naval logistics in the South Pacific and used the network model as a mixed integer linear optimization model to maximize flow.

In quantitative modeling of contested logistics, network science offers a powerful framework to model and assess the resilience of transportation systems under disruptions (Abdulla et al., 2020; Dong et al., 2020; Li et al., 2015, 2021). In this framework, logistics systems are represented as graphs composed of nodes (representing facilities, transportation hubs, and/or key infrastructure points) and edges (representing transportation links, communication channels, and/or dependencies). This abstraction allows for the application of various topological tools, such as centrality measures (e.g., betweenness and eigenvector centrality) to identify critical nodes and links whose disruption would have the most significant impact on network performance (Das et al., 2019; Duan and Lu, 2013; Haznagy et al., 2015; Reza et al., 2024; Sohouenou et al., 2020; Xie and Levinson, 2007).

For instance, studies using network science highlighted notable structural differences between road and airline networks. Road networks are typically planar (Peng et al., 2014; Xie and Levinson, 2007), with a hierarchical, mesh-like design that ensures consistent connectivity and high resilience to random disruptions (Rivera-Royero et al., 2022; Rodrigue, 2024). In contrast, airline networks often follow a scale-free topology characterized by centralized hubs and spokes. While this structure provides robustness against random failures, it leaves the network highly susceptible to targeted attacks (Barabási et al., 2000; Rodrigue, 2024; Verma et al., 2014). Understanding these structural distinctions is essential for planning, maintaining, and improving transportation infrastructure to enhance resilience and functionality.

While naval capabilities remain vital for global force projection, domestic land-based transportation via rail and highway is no less essential to military logistics. The Strategic Rail Corridor Network (STRACNET) and the Strategic Highway Network (STRAHNET), maintained by the Military Surface Deployment and Distribution Command Transportation Engineering Agency (SDDCTEA), form the backbone of the U.S. defense transportation infrastructure (SDDCTEA, 2023a, 2023b, 2022, 2013). These networks serve as vital arteries for power projection, enabling the rapid deployment of power projection platforms (PPPs), including tanks, armored vehicles, artillery, and other heavy equipment, from military installations to ports for overseas transport (FHWA, 2022). The resilience and security of STRACNET and STRAHNET are critical to ensuring the United States' ability to project military power globally and respond effectively to emerging threats.

This paper aims to quantify the resilience of land-based logistics networks by addressing the following research questions:

1. How does the vulnerability of STRACNET and STRAHNET differ under randomized versus targeted disruption scenarios regarding U.S. power projection platforms?
2. How well do the strategic transportation networks perform outside of power project platforms? Specifically, how robust and resilient are STRACNET and STRAHNET when tasked with transporting troops and equipment from any military installation to any strategic seaport?
3. What insight does the disruption modeling of transportation networks provide regarding their functional performance under large-scale disruption scenarios?

The remainder of this paper is structured as follows. We first introduce the data used to build our model, and the metrics used to characterize robustness and resilience. Next, we describe how the disruptions on the networks were modeled, firstly focusing on the specific case of the US PPPs and then considering the movements between any installations and seaports in general. Finally, we discuss key findings and limitations of this methodology, and future directions for research that can enhance network-based analyses of contested logistics.

# 2. Critical Concepts for Contested Logistics

## 2.1 Resilience vs Robustness

A traditional risk-based approach systematically identifies, assesses, and mitigates potential risks to protect and maintain critical infrastructure (CISA, 2013; ISO, 2018). In addition to risk-based approaches, two distinct concepts – *resilience* and *robustness* – are often discussed with relation to contested logistics management. *Resilience* is defined as the ability of a system to absorb shocks (*absorptivity*) and recover quickly (*rapidity*) (Bruneau and Reinhorn, 2006). In contrast to traditional risk management, which defines risk as the product of vulnerability and consequence, resilience management assumes that systems will inevitably fail and focuses on minimizing impact and improving recovery time.

*Robustness* is a related concept that describes a system's ability to withstand disruptions while maintaining functionality (Bruneau et al., 2004). While robustness focuses on preventing loss of functionality during shocks, resilience emphasizes on absorbing the initial shock and recovering rapidly afterward (Ganin et al., 2017; Omer et al., 2013; RISC, 2018). Because of these differences in their focus, robustness and resilience are both desired characteristics of critical infrastructures and they need to complement each other. A system designed to be robust against well-known and frequent disruptions may become overly specialized, leading to catastrophic failure if the intensity of a disruption exceeds expectations or if an unexpected disruption occurs. Conversely, a system that aims to be overly resilient by accounting for all potential disruptions, including frequent and predictable ones, can become inefficient due to the excessive allocation of resources for preparedness and redundancy.

## 2.2 Network Science for Contested Logistics

A network science approach to resilience analysis further integrates structural and functional characteristics to understand how systems respond to disruptions and recover from them. This methodology often employs percolation theory (Callaway et al., 2000; Deng et al., 2023; Dong et al., 2020; Li et al., 2021), which examines how the connectivity of a network evolves as nodes or links are removed due to failures or attacks. Percolation analysis identifies critical thresholds at which a network transitions from a connected state, where information or resources can flow freely, to a fragmented state, where such flow is severely hindered. By modeling networks as graphs, researchers can simulate various failure scenarios and assess their impact on overall system performance. Metrics such as connectivity, robustness, and recovery time allow researchers to quantify resilience systematically (Ganin et al., 2016; Linkov et al., 2013).

Two widely used approaches to evaluating network robustness and resilience involve *targeted* disruptions and *random* disruptions. Targeted disruptions simulate deliberate, strategic attacks on critical network components to maximize damage, while random disruptions model stochastic failures that occur unpredictably (Sohouenou et al., 2020). Although specific case studies such as the impacts of flooding in New York City (Zimmerman et al., 2023), Dayton Beach, Florida (Helderop and Grubesic, 2019), Palermo, Italy (Salvo et al., 2025), or in Norman, Oklahoma (Imran Kays et al., 2024) provide valuable localized insights, they often lack generalizability due to their context-specific nature. To address this, many studies such as (Ahmed et al., 2023, 2022; Ganin et al., 2017; Mahajan and Kim, 2020) adopt centrality-based disruption models, combining both targeted and random disruption approaches. These models offer a systematic way to evaluate network performance, identify vulnerabilities, and draw broadly applicable conclusions.

# 3. Methodology

The method and data described here closely follow those presented in (Chung et al., 2025). All network analyses were performed using Python (ver. 3.9) (See Supplemental Information).

## 3.1 Data

### *3.1.1 Highway and Rail Data*

The Strategic Highway Network (STRAHNET) data was sourced from the US Department of Transportation Bureau of Transportation Statistics as a collection of geographical lines representing interstates, non-interstate highways, and connector paths all over the US (USDOT BTS, 2024). Each road segment in the dataset included road shape and speed limits, where missing speed limits were estimated to be 55 miles per hour.

The Strategic Rail Corridor Network (STRACNET) data was sourced from the Federal Railroad Administration (FRA, 2024) and is a collection of rails that cover the continental United States and Alaska. Each rail segment in the dataset included rail geometry, but it is missing speed limits. According to the Federal Code of Regulations 49 CFR § 236.0 (c), a block signal system must be installed where freight trains are operated at speed of 50 mph or more. Since our data does not describe the existence of block signals, we used a default value of 49 mph throughout the whole STRACNET based on 49 CFR § 236.567 (c).

### *3.1.2 Military Installations and Strategic Seaports Data*

We used the publicly described military installations data depicts the locations of DoD installations (namely "forts") (USDOT BTS, 2023a). Of the 765 DoD sites across the Continental United States (CONUS), Alaska, Hawaii, and Puerto Rico, we only considered the 450 specific sites located within CONUS (Fig. 1). Each site was located with the geographic profile of the base and its military component, but only the sites located in CONUS were used in the analysis. For this analysis, no weights were given to any one site over another, except in the preliminary analysis for the U.S. power projection platforms (PPPs) described below.

The strategic seaports data includes the locations of 23 strategic seaports (namely "ports") in CONUS, Alaska, and Guam, needed to support force deployment during contingencies and other defense emergencies across the US. Among the 23 strategic seaports, we considered 21 seaports in CONUS where 17 commercial seaports were compiled by the US Maritime Administration (USDOT BTS, 2023b) and Federal Highway Administration (FHWA, 2022), and four military seaports were identified by the US Government Accountability Office (GAO, 2013).

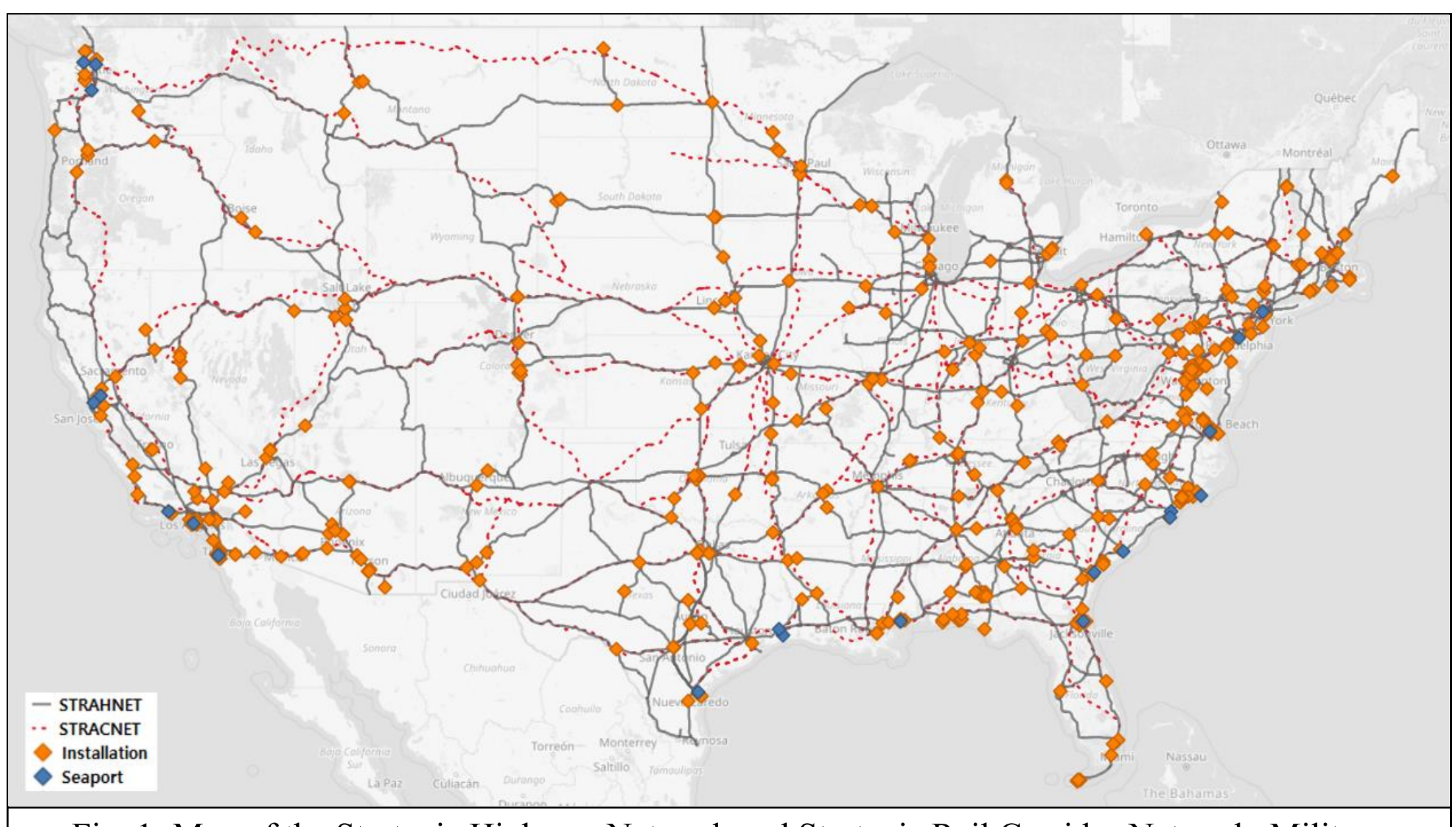

Fig. 1: Map of the Strategic Highway Network and Strategic Rail Corridor Network. Military installations and ports are distributed throughout the United States with a higher density of installations on the East and West Coasts of the United States.

The U.S. power projection platforms (PPPs) are key military installations strategically positioned to facilitate the rapid deployment of forces and equipment (FHWA, 2022). As of 2022, there are 17 listed PPPs consisting of 15 US Army Installations (i.e. Forts) and two US Marine Corps installations (i.e. Camps) critical to US power projection (Table 1). Each Army and Marine Corps PPP is associated with a designated seaport of embarkation (SPOE, hereinafter Seaport) and aerial port of embarkation (APOE, hereinafter Airport) for overseas deployment. While some installations have seaports in proximity, others rely on highway and rail infrastructure as a primary mode of transport to connect with more distant seaports. Note that some airports are co-located with the main installations (E.g. Fort 16 and Airport 16). The PPP installation names, their corresponding ports of embarkation, and the distances and travel times between them have been anonymized and discretized in this paper to protect potentially sensitive information (Tables 1–3).

### 3.2 Transportation Graph

The backbone of military logistics is transportation networks. To model and analyze the transportation networks, we represent the physical network by a mathematical graph with well-defined nodes and edges. In our graph representation of highway and railway networks, the nodes represent road/rail intersections, and the edges represent road/ rail segments connecting the nodes. GIS based refinements as outlined in (Chung et al., 2025) were implemented to align our model with the actual topology and connectivity of the physical highway and rail networks. A detailed description of the data cleaning tasks can be found in Supplemental Information.

The transportation networks were modeled as undirected graphs, with each edge assigned a travel time calculated by dividing its length by its speed limit. In this representation, all trips must begin and end at nodes, and transitions between road segments, such as turning, can only occur at nodes. To integrate military installations and seaports into the transportation graph, each facility was mapped to the geometrically closest point on the graph. If the closest point lies in the middle of an edge, that edge was split into two segments, and a new node was inserted at the split location.

### 3.3 Performance Measures

#### *3.3.1 Origin-Destination Travel Time*

For each origin-destination pair ($od$ pair), the shortest path can be determined by the Dijkstra's algorithm (Dijkstra, 1959) minimizing the total sum of edge travel times. The travel time matrix $T$ is a $|O| \times |D|$ dimensional matrix representing travel times along the shortest paths for all possible $od$ paths. Each element $t_{od}$ of $T$ represents the travel time between an origin $o \in O$ and a destination $d \in D$. Since the undisrupted transportation network $G$ is connected and $o \neq d$ for any $od$ pairs, the travel time value $t_{od}$ is always positive and finite.

For the analysis of PPPs, only the routes between each installation to its designated seaport and airport were considered. In the generalized all forts to all ports scenario, we analyzed all combinatorial pairs for accessible installations, where 9000 possible fort-to-port paths (450 forts, 20 ports) using the highways and 4675 paths (275 forts, 17 ports) using the railways were possible.

Table 1: Specifications of the US Power Projection Platforms

| PPP Installation | To Seaport | | | | | To Airport | | | | |
|---|---|---|---|---|---|---|---|---|---|---|
| | SPOE | Highway | | Railway | | APOE | Highway | | Railway | |
| | | Length (km) | Travel Time (h) | Length (km) | Travel Time (h) | | Length (km) | Travel Time (h) | Length (km) | Travel Time (h) |
| Fort 1 | Seaport 1 | 500-1000 | 5 | 500-1000 | 8 | Airport 1 | on-site | | | |
| Fort 2 | Seaport 1 | 0-500 | 1 | 0-500 | 1 | Airport 2 | on-site | | | |
| Fort 3 | Seaport 2 | 1500-2000 | 18 | 2000-2500 | 26 | Airport 3 | 0-50 | 0 | 0-50 | 1 |
| Fort 4 | Seaport 3 | 0-500 | 0 | 0-500 | 4 | Airport 4 | on-site | | | |
| Fort 5 | Seaport 4 | 0-500 | 1 | 0-500 | 1 | Airport 5 | 50-100 | 1 | 0-50 | 1 |
| Fort 6 | Seaport 5 | 0-500 | 2 | 0-500 | 3 | Airport 6 | on-site | | | |
| Fort 7 | Seaport 6 | 0-500 | 1 | 0-500 | 1 | Airport 7 | 50-100 | 1 | 50-100 | 1 |
| Fort 8 | Seaport 6 | 0-500 | 4 | 0-500 | 6 | Airport 8 | on-site | | | |
| Fort 9 | Seaport 7 | 1000-1500 | 11 | 1000-1500 | 16 | Airport 9 | on-site | | | |
| Fort 10 | Seaport 8 | 1000-1500 | 14 | 1500-2000 | 25 | Airport 10 | 100-200 | 1 | 100-200 | 1 |
| Fort 11 | Seaport 8 | 500-1000 | 8 | 1000-1500 | 17 | Airport 11 | on-site | | | |
| Fort 12 | Seaport 8 | 500-1000 | 6 | 500-1000 | 8 | Airport 12 | 50-100 | 1 | 800-900 | 12 |
| Fort 13 | Seaport 8 | 0-500 | 4 | 500-1000 | 9 | Airport 13 | on-site | | | |
| Fort 14 | Seaport 8 | 1000-1500 | 9 | 1500-2000 | 22 | Airport 14 | on-site | | | |
| Fort 15 | Seaport 8 | 1500-2000 | 17 | 2000-2500 | 26 | Airport 15 | 0-50 | 0 | 0-50 | 0 |
| Fort 16 | Seaport 9 | 0-500 | 2 | 500-1000 | 10 | Airport 16 | on-site | | | |
| Fort 17 | Seaport 10 | 0-500 | 1 | 0-500 | 1 | Airport 17 | 100-200 | 2 | 100-200 | 2 |
| | Average | 655 | 6.2 | 847 | 10.8 | Average | 79 | 1.3 | 187 | 2.4 |
| | % Network Utilized | 10.90% | | 20.90% | | % Network Utilized | 0.50% | | 1.90% | |

### 3.3.2 *Edge betweenness centrality*

Edge betweenness centrality is a metric measures shortest-path centrality that has been widely adopted across various domains of network science to analyze connectivity, flow, and system vulnerabilities which quantifies the importance of an edge within a network (Freeman, 1977). It is defined as the "number of shortest paths that pass through an edge in a graph" (Lu and Zhang, 2013), capturing its role as a critical link in maintaining network connectivity and flow. Mathematically, the betweenness centrality of an edge $e$ is expressed as:

$$c_B(e) = \sum_{o \in O, d \in D} \frac{\sigma(o, d|e)}{\sigma(o, d)}$$

where $O$ represents the set of origins, $D$ denotes the set of destinations, $\sigma(o, d)$ is the total number of shortest paths between nodes $o$ and $d$, and $\sigma(o, d|e)$ is the number of those paths that pass through edge $e$ (Brandes, 2008).

## 3.4 Disruptions

Disruptions are incidents that prevent systems from functioning properly. In our case, disruption is represented by road or rail segment removal from a transportation network and its impact on logistics is quantified by the resulting time delay on fort-to-port paths. Many natural disasters such as floods or wildfires, and artificial disasters such as car crashes or terrorism can disrupt transportation networks.

### *3.4.1 Disruptions for Power Projection Platforms*

Since there are only 17 fort-to-seaport paths and 7 fort-to-airport paths in PPPs which utilizes only about 11 % of the highway network and 23 % of the rail network (Table 1), we focus on disrupting individual paths rather than the disruptions in the whole transportation network. We analyzed the impact of instantaneous short disruptions which may impact the ability for PPPs to quickly project power across the globe. For each fort-to-port PPP path, we measured the impact of removal of each edge along the path. Here, the road length of the removed edge is considered as the cost of the disruption and the travel time delay caused by the removal of an edge measures the importance of the edge.

### *3.4.2 Disruptions for General Fort-to-Port Movements*

In contrast to the PPP model which focuses only on 24 specific paths, there are thousands of fort-to-port paths that need to be considered in the general case which. To analyze the robustness and resilience of the entire transportation network, we modeled two types of disruptions, namely *targeted* and *random*, with varying degrees of intensity.

The *intensity* of a disruption is measured in the percentage of road or rail lengths that were removed. For our simulations, disruption intensities ranged from 1 % to 50 % with a 1 % increment. Quantifying disruption intensities based on the total length of affected roads and rails enables us to compare the effects of different disruptions across various transportation modes.

The *targeted* disruption removes roads and rail segments based on their importance, measured by their betweenness centrality. Under a targeted disruption, roads and rails with the highest centrality values are removed first, until the percentage of the total road and rail lengths removed equals the desired intensity of the disruption. Because the centrality of the roads was determined prior to the disruptions, the targeted disruptions are deterministic.

On the other hand, under a *random* disruption, road and rail segments are removed at random such that the total percentage of lengths removed equals the desired intensity. Since the random disruptions are stochastic, we repeat random disruptions 100 times for each intensity to observe their statistical behaviors. Examples of targeted and disrupted target profiles can be seen in Supplemental Information.

### *3.4.3 Delay Characterization*

To measure the impact of disruptions, travel times before a disruption $t_{od}$ and after a disruption $t'_{od}$ were collected, for each $od$ pair. These values satisfy $0 < t_{od} < \infty$ and $0 < t'_{od} \leq \infty$ such that $t_{od} \leq t'_{od}$ where $t'_{od} = \infty$ indicates disconnection from an origin $o$ to a destination $d$.

By measuring the time increase due to a disruption, we can quantify the impact of the disruption on each path. Here, we use the relative time difference $\Delta t_{od} \coloneqq \frac{t'_{od} - t_{od}}{t_{od}}$ measured in percentage time increase with respect to the undisrupted travel time. Using $\Delta t_{od}$, we classify the impact of the disruption on each path into three categories where the path $od$ is *unaffected* if $\Delta t_{od} = 0$, *delayed* if $0 < \Delta t_{od} < \infty$, and *disconnected* if $\Delta t_{od} = \infty$.

This trichotomy of paths satisfy the relation $N = N_{unaffected} + N_{delayed} + N_{disconnected}$, where $N = |O| \cdot |D|$ is the total number of paths, and $N_{unaffected}$, $N_{delayed}$, and $N_{disconnected}$ represent the number of unaffected, delayed, and disconnected paths, respectively.

### 3.5 Impacts of Disruptions

#### *3.5.1 Disrupted U.S. Power Projection Platforms*

For the analysis on U.S. power projection platforms, we analyzed the impact of single point disruptions which may impact the ability for PPPs to quickly project power across the globe. Disruptions in the PPP model were single point disruptions of disabling one edge $e$ on the shortest path between origin and destination pair $od$. The $N$ edges on the shortest path from origin $o$ to destination $d$ were ordered by their respective delay values $\Delta t_1, \Delta t_2, \ldots, \Delta t_N$, each associated with an edge length $l_1, l_2, \ldots, l_N$ representing the cost of disruption. The 50th percentile delay, calculated as a length-weighted median, was defined as the $k^{\text{th}}$ delay where $k$ satisfies

$$\sum_{i=1}^{k-1} \frac{l_i}{L} \leq \frac{1}{2} \quad \text{and} \quad \sum_{i=k+1}^{n} \frac{l_i}{L} \leq \frac{1}{2}$$

where $L = \sum_{i=1}^{N} l_i$ is the total length of the route. The 50th percentile disruption represents the disruption intensity at which a “randomly” placed failure is equally likely to occur before or after this point along the known shortest path. In addition to the 50th percentile delay, the 25th and 75th percentiles and the maximum disruption for each installation and port pair were also calculated (Table 2 and 3).

#### *3.5.2 Disrupted Strategic Transportation Networks*

Since the U.S. PPP analysis focuses on a small set of specific routes, the result may not be reflecting the structure of the whole transportation networks. To analyze the transportation network as a whole, an aggregate measure considering the impact of disruptions on all possible fort-to-port paths need to be defined. Previous percolation analyses (Callaway et al., 2000; Deng et al., 2023; Dong et al., 2020; Li et al., 2021) focused on network connectivity and the giant connected component, where the paths in the network are either connected or disconnected. Because time delays $\Delta t_{od}$ are continuous values, connectivity-based analyses may underestimate the impact of disruptions. To include the continuous characteristics of time delays, we define a *functionality* function on paths that calculates how ‘functional’ the path from $o$ to $d$ is with respect to the undisrupted network. We calculated the functionality of a delayed path as the monotonically decreasing function $f(\Delta t) = \frac{1}{1+\Delta t}$. Thus, our *functionality* measure $f(\Delta t_{od}) \in [0,1]$ indicates that the path $od$ is:

1) fully functional and unaffected by the disruption if $f(\Delta t_{od}) = 1$,
2) completely disconnected by the disruption if $f(\Delta t_{od}) = 0$, and
3) delayed by the disruption if $0 < f(\Delta t_{od}) < 1$.

Finally, we define the functionality of the network under a disruption as the average functionality of individual fort-to-port paths under the disruption: $f(G) := \frac{1}{N} \sum_{o,d} f(\Delta t_{od})$.

## 4. Results

### 4.1 Disrupted U.S. Power Projection Platforms

The shortest paths specification between each PPP installation and their associated Seaport and Airport are reported for the highway (Table 2) and railway networks (Table 3). These shortest paths constituted approximately 11 % of the highway network and 23 % of the rail network (Table 1). This finding indicates that for the major mission of power-projection, a majority of railway and highway lines do not play a critical role.

#### *4.1.1 Installation to Seaport*

In general, the path between each PPP installation and its associated seaport was shorter using highways than railways. The mean travel distance was 655 km by highway and 847 km by railway. Five installations, including both Marine Corps bases, had the shortest highway paths to seaports that were under 1 hour. Meanwhile, six installations had travel distances over 1000 km by rail. For all installations, highways represented a quicker path to seaports than those of railways.

The 50th percentile disruption to highway networks translated to an average of a 1 hour increase in travel time which corresponds to an average of a 54.6 % increase in travel times. Seven installations had less than 10 % increases in travel times at the 50th percentile. For most installation-seaport pairs, the highway network was relatively robust against random single point disruptions, with only three seaport-installation pairs having greater than 100 % of time delays associated with a 50th percentile randomized disruptions (Table 2).

By contrast, the maximum impact of single point disruptions caused complete disconnection in all cases except one (i.e. Base3 to Seaport3). These relatively high impacts of targeted disruptions compared to random disruptions suggest that most random disruptions have relatively limited impacts on the travel time requirement for most installations. This suggests that U.S. PPPs using highways are highly vulnerable to targeted single point disruptions, but robust against random disruptions.

On the other hand, both the 50th percentile and maximum single point disruptions were associated with full disconnection for multiple installation-port pairs. For 50th percentile disruptions that did not fully disconnect railway networks, 50th percentile disruptions were associated with a 236 % increase in travel times. Only two installations had less than 10 % increases in travel times at the 50th percentile. The maximal disruptions for railways were always disconnecting the installations to seaports and more than half of the paths had at least 1000 % delay or even full disconnections even at the 75th percentile level.

Additionally, we found that regardless of the transportation mode, the road and rail segments that causes maximal disruptions are located at the beginning near the installation and at the end near the ports of embarkation (Supplemental Information 2.2). When using highways to reach designated ports of embarkation, the total length of roads that cause maximal disruption was at most 30 km in most of PPPs. This result is different for railways where the total length of rails that cause maximal disruption was at least 100 km in most cases.

Table 2: Impact of Highway Disruptions on Shortest Paths between Installations and Ports of Embarkation in the US Power Projection Platforms.

| **PPP Installation** | **Port of Embarkation** | **Highway** | | | | | | | | | |
|---|---|---|---|---|---|---|---|---|---|---|---|
| | | **25th percentile** | | **50th percentile** | | **75th percentile** | | **maximum disruption** | | | |
| | | time increase (h) | % time increase | time increase (h) | % time increase | time increase (h) | % time increase | time increase (h) | % time increase | length causing maximum (km) | % length causing maximum |
| Seaport of Embarkation | | | | | | | | | | | |
| Fort 1 | Seaport 1 | 1 | 11.2% | 1 | 17.0% | 1 | 17.0% | inf | inf | 100-200 | 21.7% |
| Fort 2 | Seaport 1 | 1 | 55.0% | 1 | 65.2% | 1 | 65.2% | inf | inf | 0-100 | 14.0% |
| Fort 3 | Seaport 2 | 0 | 1.5% | 1 | 3.0% | 1 | 3.5% | 5 | 26.6% | 0-100 | 1.5% |
| Fort 4 | Seaport 3 | 0 | 29.4% | 0 | 29.4% | 0 | 29.4% | inf | inf | 0-100 | 8.5% |
| Fort 5 | Seaport 4 | 1 | 133.8% | 1 | 133.8% | 1 | 133.8% | inf | inf | 0-100 | 15.9% |
| Fort 6 | Seaport 5 | 0 | 2.5% | 0 | 2.5% | 0 | 8.1% | inf | inf | 0-100 | 0.6% |
| Fort 7 | Seaport 6 | 0 | 13.6% | 0 | 13.6% | 0 | 29.2% | inf | inf | 0-100 | 22.8% |
| Fort 8 | Seaport 6 | 0 | 2.7% | 1 | 12.5% | 1 | 27.1% | inf | inf | 0-100 | 4.0% |
| Fort 9 | Seaport 7 | 0 | 0.7% | 0 | 0.7% | 0 | 0.7% | inf | inf | 0-100 | 0.5% |
| Fort 10 | Seaport 8 | 0 | 0.8% | 1 | 4.6% | 1 | 5.9% | inf | inf | 0-100 | 2.0% |
| Fort 11 | Seaport 8 | 0 | 0.2% | 1 | 7.5% | 1 | 8.8% | inf | inf | 0-100 | 2.6% |
| Fort 12 | Seaport 8 | 1 | 21.3% | 1 | 21.3% | 2 | 37.6% | inf | inf | 0-100 | 3.7% |
| Fort 13 | Seaport 8 | 0 | 0.3% | 0 | 0.3% | 0 | 2.3% | inf | inf | 0-100 | 5.3% |
| Fort 14 | Seaport 8 | 1 | 13.0% | 1 | 13.0% | 1 | 13.0% | inf | inf | 0-100 | 2.2% |
| Fort 15 | Seaport 8 | 1 | 4.0% | 1 | 4.1% | 1 | 6.9% | inf | inf | 0-100 | 1.3% |
| Fort 16 | Seaport 9 | 4 | 169.0% | 4 | 169.0% | 4 | 169.0% | inf | inf | 0-100 | 2.3% |
| Fort 17 | Seaport 10 | 0 | 12.6% | 3 | 431.5% | 3 | 431.5% | inf | inf | 0-100 | 0.9% |
| Seaport Average | | 6.1 | 27.7% | 0.9 | 54.6% | 1.1 | 58.2% | 4.9 | 26.6% | 21.2 | 6.5% |
| Airport of Embarkation | | | | | | | | | | | |
| Fort 3 | Airport 3 | 2 | 525.3% | 6 | 1408.8% | 6 | 1408.8% | 6 | 1408.8% | 0-100 | 63.7% |
| Fort 5 | Airport 5 | 0 | 43.1% | 0 | 43.1% | 0 | 43.1% | inf | inf | 0-100 | 5.4% |
| Fort 7 | Airport 7 | 0 | 16.9% | 0 | 16.9% | 0 | 67.8% | inf | inf | 0-100 | 19.7% |
| Fort 10 | Airport 10 | 3 | 331.4% | 3 | 331.4% | 3 | 331.4% | inf | inf | 0-100 | 4.5% |
| Fort 12 | Airport 12 | 3 | 348.3% | 3 | 348.3% | 3 | 348.3% | inf | inf | 0-100 | 0.2% |
| Fort 15 | Airport 15 | 3 | 1761.7% | 3 | 1761.7% | 19 | 10151.6% | inf | inf | 0-100 | 3.8% |
| Fort 17 | Airport 17 | 0 | 15.3% | 0 | 15.3% | 0 | 15.3% | 0 | 15.3% | 100-200 | 90.8% |
| Airport Average | | 1.8 | 434.6% | 2.3 | 560.8% | 4.6 | 1766.6% | 3.1 | 712.1% | 31.5 | 27% |

### *4.1.2 Installation to Airport*

Unlike paths to seaports which often are hundreds of kilometers away from the installations, paths to airports are relatively close and even co-located with the installation. 10 out of 17 installations had their airfields on-site and, even those with off-site airports, most of them could be reached within two hours.

When using highways, three out of seven paths had less than 1 hour of delay, and the remaining four paths had delays of more than 3 hours. Against maximal disruptions, most of them were disconnected or had 1400 % increase in travel time. The only exception was the route between Base17 and Airport17, which benefited from the relatively dense highway infrastructure. On the other hand, when using railways, even the 25th percentile disruptions significantly increased travel times by over 8 hours or fully disconnected.

Table 3: Impact of Railway Disruptions on Shortest Paths between Installations and Ports of Embarkation in the US Power Projection Platforms.

| PPP Installation | POE | Railway | | | | | | | | | |
|---|---|---|---|---|---|---|---|---|---|---|---|
| | | 25th percentile | | 50th percentile | | 75th percentile | | maximum disruption | | | |
| | | time increase (h) | % time increase | time increase (h) | % time increase | time increase (h) | % time increase | time increase (h) | % time increase | length causing maximum (km) | % length causing maximum |
| Seaport of Embarkation | | | | | | | | | | | |
| Fort 1 | Seaport 1 | 5 | 62% | 6 | 77% | 6 | 77% | inf | inf | 100-200 | 21.2% |
| Fort 2 | Seaport 1 | 3 | 199% | 11 | 792% | 11 | 792% | inf | inf | 0-100 | 12.6% |
| Fort 3 | Seaport 2 | 1 | 5% | 1 | 5% | 9 | 33% | inf | inf | 100-200 | 6.3% |
| Fort 4 | Seaport 3 | inf | inf | inf | inf | inf | inf | inf | inf | 200-300 | 81.0% |
| Fort 5 | Seaport 4 | inf | inf | inf | inf | inf | inf | inf | inf | 0-100 | 99.5% |
| Fort 6 | Seaport 5 | 0 | 0% | inf | inf | inf | inf | inf | inf | 100-200 | 72.9% |
| Fort 7 | Seaport 6 | 9 | 1044% | 9 | 1044% | 9 | 1044% | inf | inf | 0-100 | 14.6% |
| Fort 8 | Seaport 6 | 0 | 7% | 0 | 7% | 0 | 7% | inf | inf | 0-100 | 13.4% |
| Fort 9 | Seaport 7 | 1 | 8% | 3 | 22% | 4 | 23% | inf | inf | 0-100 | 2.2% |
| Fort 10 | Seaport 8 | 4 | 15% | 4 | 15% | 12 | 47% | inf | inf | 300-400 | 18.8% |
| Fort 11 | Seaport 8 | 0 | 0% | 4 | 21% | inf | inf | inf | inf | 300-400 | 27.7% |
| Fort 12 | Seaport 8 | inf | inf | inf | inf | inf | inf | inf | inf | 500-600 | 79.2% |
| Fort 13 | Seaport 8 | 4 | 40% | inf | inf | inf | inf | inf | inf | 400-500 | 57.6% |
| Fort 14 | Seaport 8 | 2 | 11% | 2 | 11% | 16 | 73% | inf | inf | 300-400 | 22.2% |
| Fort 15 | Seaport 8 | 4 | 16% | 12 | 45% | 13 | 50% | inf | inf | 300-400 | 18.5% |
| Fort 16 | Seaport 9 | 41 | 430% | 53 | 556% | inf | inf | inf | inf | 200-300 | 29.8% |
| Fort 17 | Seaport 10 | inf | inf | inf | inf | inf | inf | inf | inf | 100-200 | 100.0% |
| Seaport Average | | 5.7 | 141.3% | 9.6 | 235.8% | 8.8 | 238.4% | inf | inf | 209.7 | 39.9% |
| Airport of Embarkation | | | | | | | | | | | |
| Fort 3 | Airport 3 | 18 | 3581.3% | 18 | 3581.3% | 18 | 3581.3% | 18 | 3581.3% | 0-100 | 100% |
| Fort 5 | Airport 5 | inf | inf | inf | inf | inf | inf | inf | inf | 0-100 | 98.7% |
| Fort 7 | Airport 7 | 9 | 1140.0% | 9 | 1140.0% | 9 | 1140.0% | inf | inf | 0-100 | 15.7% |
| Fort 10 | Airport 10 | 24 | 1737.9% | 24 | 1737.9% | 24 | 1737.9% | 24 | 1737.9% | 100-200 | 92.9% |
| Fort 12 | Airport 12 | 8 | 66.8% | 8 | 66.8% | 8 | 66.8% | inf | inf | 100-200 | 16.5% |
| Fort 15 | Airport 15 | 34 | 16871.2% | 34 | 16871.2% | 34 | 16871.2% | 34 | 16871.2% | 0-100 | 94.3% |
| Fort 17 | Airport 17 | 1 | 72.5% | inf | inf | inf | inf | inf | inf | 0-100 | 66.8% |
| Airport Average | | 15.6 | 3911.6% | 18.5 | 4679.4% | 18.5 | 4679.4% | 25.1 | 7396.8% | 64.9 | 69.3% |

## 4.2 Disrupted U.S. Strategic Transportation Networks

When considering all possible fort-to-port paths across CONUS, the transportation network's structure becomes more important since the installations are distributed across the entire CONUS. Because strategic seaports are distributed primarily on the East and West Coast, cross-continental east–west interstates (e.g. I-10, I-20, I-80, I-90) and similarly oriented railways (e.g. Union Pacific Lines in the Southwest) had higher centrality values (See Supplemental 2.1).

Using the delay characterization, both random and targeted disruptions on both the highway and rail network share similar degradation profiles as the intensity of disruptions increases (Fig. 2). When disruption intensity is very small (less than 3 %), we have an absorptive phase where delayed routes increase, but disconnections have not occurred yet. When the disruption intensity increases, we have a high-delay phase where the number of delayed paths does not change much around 60 %, but the number of disconnected paths increases. When the network is under severe disruptions, we have the dismantle phase where delayed paths become disconnected and the network is broken down into smaller connected components.

Even though both highways and railways share similar qualitative degradation profile, highways have higher functionality value than the railways under the same type of disruption (Fig. 3). The highway network was 70 % functional under 10 % targeted disruption, whereas railways already lost significant functionality and are only 30 % functional. Similarly, under 10 % random disruption, highways are on average 65 % functional but railways are only 30 % functional on average.

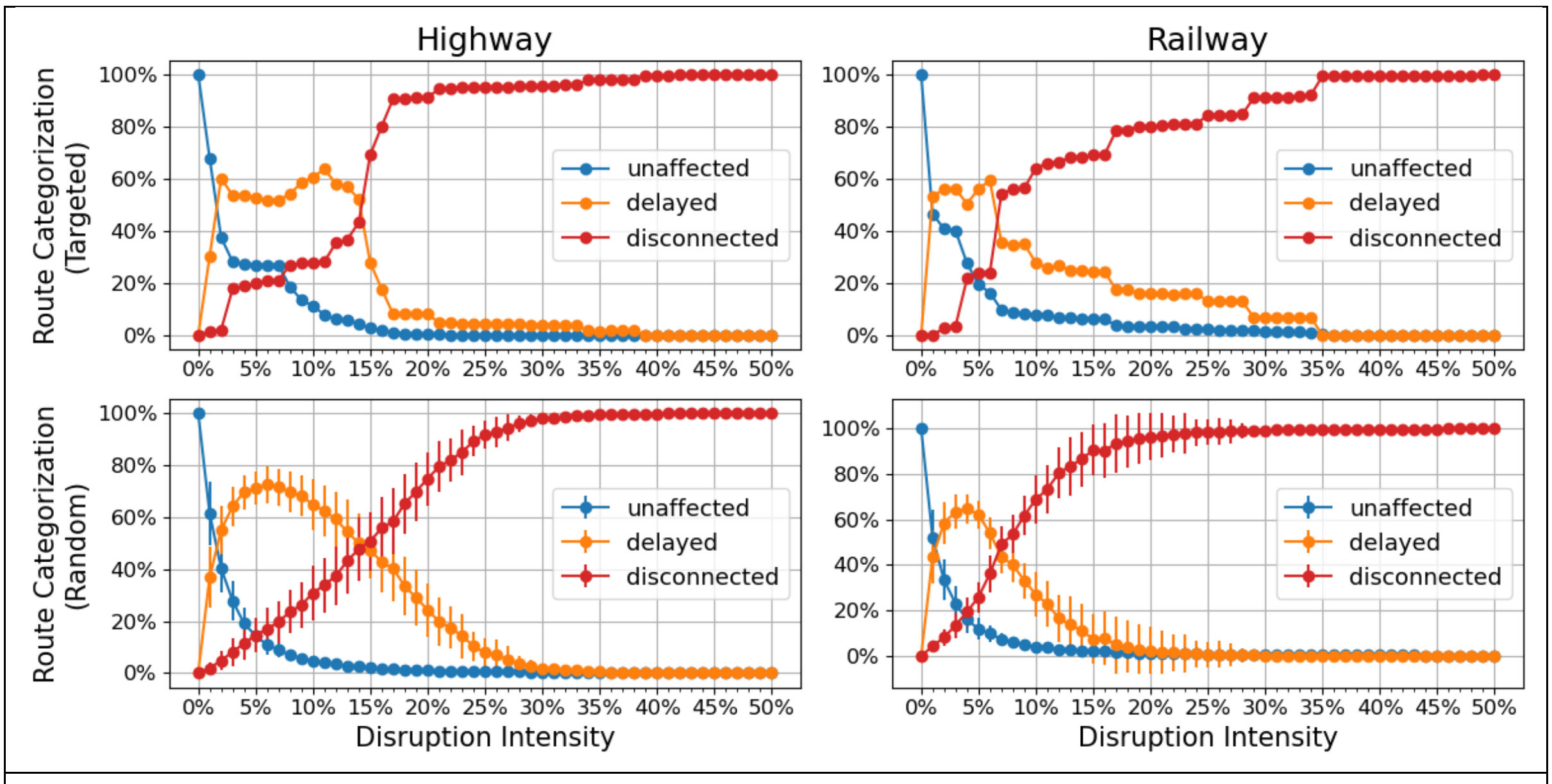

Fig. 2: Impacts of the disruption based on delay characterization for the targeted (Top) and random (Bottom) attacks. The proportion of unaffected nodes associated with a disruption decreases as intensity increases

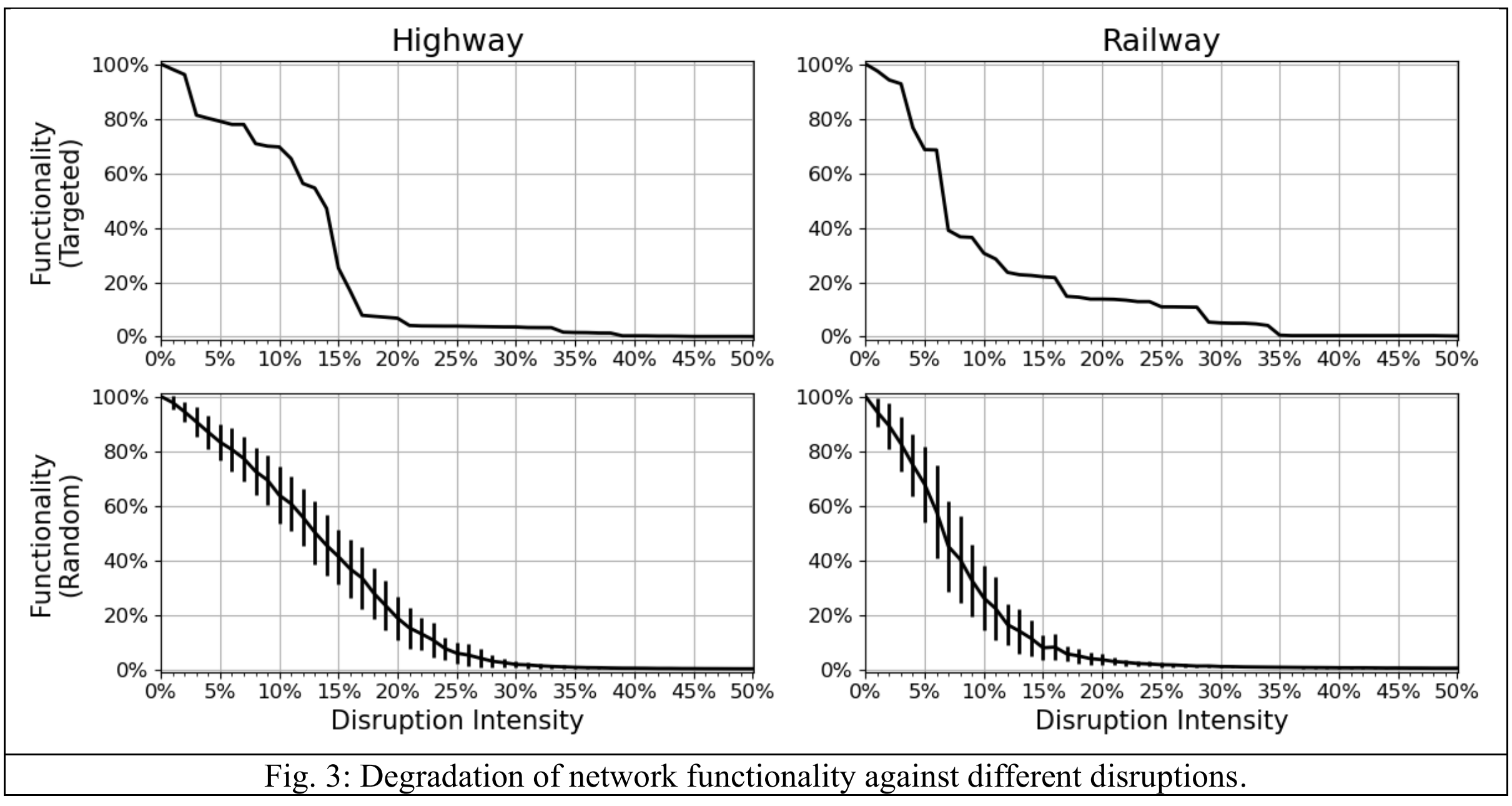

Fig. 3: Degradation of network functionality against different disruptions.

# 5. Discussions

In this paper, we developed a simplified methodology to characterize contested logistics both under short acute disruptions as well as major disruptions that disable large portions of the logistics network. We found that for specific origin–destination pairs based on PPPs, small sections of the highway network could be protected to ensure connectivity. Most critical disruptions only occurred at the “first-mile” or “last-mile” of the route. Thus, protecting against disruptions in the road network may not just include road hardening, but planning alternative routes closer to the installation and ports. However, based on the total lengths of roads and rails that can cause maximal disruptions, it might be more difficult to strengthen railways since it would require hardening and building detours of length greater than 100 km in many cases, whereas less than 30 km is required for highways.

We also found that different target-prioritization methods can have key differences in the impact on the road network. We found that the critical degradation of the network is more pronounced under targeted disruption scenario compared to the random disruption scenario, especially after a certain disruption intensity threshold (around 15 % in our model). This suggests that understanding potential adversary targeting strategies and prioritizing defense efforts accordingly is critical to understanding how a system will degrade over an adverse event. The edge betweenness centrality metric proved useful in identifying these critical links, but more sophisticated models could incorporate factors like asset value, repair time, and strategic importance to refine targeting priorities.

Several abstractions were made to simplify the model for general understandability. Notably, the targeting mechanism does not include any, nor was it optimized for sequential disruptions or the true cost of disrupting an edge. Methods to completely disable a road or rail- network are likely to be much more time and resource intensive. Roads and rail can be rapidly restored to meet mission needs, especially during critical moments, if anticipated properly. While the cost of disabling an edge was based on the length of a given edge, under the assumption that actually causing a disruption to that edge would require disrupting multiple parts of that road or railway to ensure it could not be used. However, single point disruptions could potentially stall the succession of disruptions. Examples of this can be seen in the case of the Russian invasion of Ukraine where poor road conditions may have caused convoys to fail. Network model tools such as the one presented in this paper can be used to identify the most vulnerable segments and design strategies to mitigate these “single point of failure”.

Mixed-mode analysis could potentially dramatically shift the risk landscape. This analysis only characterized either the highway or rail- way network, but real-world logistics systems often involve a combi- nation of transportation modes. A more comprehensive analysis would integrate both highway and railway networks, as well as air and sea transport, to assess the resilience of end-to-end supply chains. However, in the contested logistics environment, this will require the analysis of loading and unloading of highly specialized equipment that cannot easily be captured by network analysis. The ability to switch between modes in response to disruptions could significantly enhance overall system resilience. Further, by considering the interdependence between these modes (e.g., rail lines feeding ports), we can identify critical choke points that require prioritized protection.

Even though we suggested a new GIS and network science-based method to identify vulnerabilities and to measure robustness and resilience of logistics networks under various disruption scenarios, more work can be done to improve our findings. Firstly, even though the recovery phase is a crucial part of measuring resilience, it has not been modeled and included here yet. Once the disruption peaks, one can model various recovery strategies to optimize for rapid recovery of the system and suggest effective recovery plans. By using more realistic logistics models and disruption models, our contested logistics model would better reflect the real-world system..

## 6. Conclusions

This paper presented a simplified methodology for quantifying resilience in contested logistics environments, focusing on the US STRACNET and STRAHNET. By leveraging network science principles and readily available data, we characterized the vulnerability of these critical land networks to both random and targeted disruption scenarios. Our findings revealed differences in the resilience of highway and rail-way systems, specifically in the degradation profile of networks. This analysis highlighted key vulnerabilities in last-mile components of the US power projection platforms and demonstrated the impact of targeted disruptions on overall network functionality. While the model in- corporates necessary abstractions, it provides a valuable framework for understanding the complex interplay between network structure, disruption intensity, and system performance. The analysis underscores the importance of strategic infrastructure hardening, alternative route planning, and a comprehensive understanding of potential adversary targeting strategies. Furthermore, the discussion of limitations and proposed future research directions lays a roadmap for developing more sophisticated and realistic models that incorporate dynamic network conditions, repair and recovery processes, cost-benefit considerations, and multi-modal transportation integration. Ultimately, this research contributes to a growing body of knowledge aimed at enhancing the resilience of critical supply chains, ensuring the ability to project power, and maintaining economic stability in an increasingly contested and uncertain global landscape.

## Acknowledgements

The authors declared no potential conflicts of interest with respect to the research, authorship, and/or publication of this article.

During the preparation of this work the authors used ChatGPT and Perplexity in order to enhance readability of the manuscript. After using this tool/service, the authors reviewed and edited the content as needed and take full responsibility for the content of the publication.